\documentclass[prb,twocolumn,superscriptaddress,showpacs]{revtex4-1}
\usepackage{epsfig}
\usepackage{dcolumn}
\usepackage{epstopdf}
\usepackage{graphicx}
\usepackage{color}
\begin{document}
\title{Large band offset as driving force of 2-dimensional electron confinement: the case of SrTiO$_3$/SrZrO$_3$ interface}
\author{P. Delugas}\author{A. Filippetti}
\affiliation{CNR-IOM, UOS Cagliari, S.P. Monserrato-Sestu Km. 0.700, Monserrato (CA), Italy}
\author{A. Gadaleta}\author{I. Pallecchi}\author{D. Marr\'e}
\affiliation{CNR-SPIN UOS Genova and Dipartimento di Fisica, Via Dodecaneso 33, 16146 Genova, Italy}
\author{V. Fiorentini}
\affiliation{CNR-IOM, UOS Cagliari and Dipartimento di Fisica, Universit\`a di Cagliari, Monserrato (CA), Italy}
\date{\today}
\begin{abstract}
Using advanced first-principles calculations we predict that the non-polar SrTiO$_3$/SrZrO$_3$ (001) interface, designed as either thin SrZrO$_3$ film deposited on SrTiO$_3$ or short-period (SrTiO$_3$)$_m$/(SrZrO$_3$)$_n$ superlattice, host a 2-dimensionally confined electron gas. Mobile electron charge due to native impurities, field-effect, or modulation doping remains tightly trapped at the interface. Key ingredients for this occurrence are a) the peculiar chemistry of 3d orbitals, b) the large band offset at titanate-zirconate interface.
\end{abstract}

\maketitle

\section{Introduction}

The observations of a 2-dimensional electron gas (2DEG) in SrTiO$_3$/LaAlO$_3$ (STO/LAO) interfaces and superlattices\cite{ohtomo,thiel,huijben,reyren,nakagawa,herranz} triggered a gold rush toward the discovery and design of similar oxide heterostructures that would open the way to a new all-oxide nanoelectronics. Despite extensive efforts, however, 2DEGs in oxide heterostructures remain rare, and a consensus about genuine 2DEG formation emerged only for a handful of systems other than STO/LAO. Among single interfaces we can mention the n-type STO/LaVO$_3$\cite{hotta}, STO/LaGaO$_3$\cite{perna}, STO/GaTiO$_3$\cite{son,moe,cain,boucherit}, and some amorphous STO, LAO, and Al$_2$O$_3$ overlayers grown on STO\cite{chen,lee}. Interestingly, a 2DEG was also revealed at the (001) surface of STO thin-films\cite{syro} and at the (001) surface of KTaO$_3$\cite{king,syro1}. An alternative route to the spontaneous 2DEG formation is based on the explicit inclusion of doping layers, e.g. single Nb-doped TiO$_2$ layers introduced in epitaxial STO films in form of a superlattice (SL)\cite{ohta_nm,mune,ohta_tsf}, and single RO layers (with R=La, Pr, Nd, Sm) substitutional to SrO in STO\cite{jang}.

From the conceptual perspective, years of studies were not yet sufficient to reach, from this variety of observations, a common understanding on the essential ingredients causing the 2DEG formation in oxide heterostructures. The polar character of the overlayer, the correlated nature of Ti 3d electronic states, the large band-offset at the interface, the dielectric properties of the STO substrate are all generically invoked as 2DEG driving forces, but without a clear definition of what is really essential and what it is not. In particular, one of the most debated aspects concerns the role of film polarity and the related polarization catastrophe. The recent observation of 2DEG at the non-polar (110) STO/LAO interface\cite{herranz} suggests that polarity is in fact not so essential to the 2DEG presence. Theory and simulation, especially based on ab-initio methods, can give an invaluable contributions to this conceptual analysis, not only for their typical accuracy, but also for the capability to enlighten the instrinsic behaviour of the ideal systems (i.e. filtered by the complicacies introduced by disorder, inhomogeneities, structural imperfections, etc.).  

Following this line of thought, in this work we furnish a sound proof of concept concerning the presence of 2DEG in an ideal (disorder and defect-free) non-polar oxide heterostructures in form of both single interface and superlattice. Specifically, we predict that a 2DEG system tightly confined to a few atomic layers occurs at the (001) interface between SrTiO$_3$ (STO) and SrZrO$_3$ (SZO)\cite{shafranek,kajdos}. Unlike STO/LAO, neither side of the STO/SZO interface is polar, hence no polarization catastrophe can occur and the system is intrinsically insulating. Nevertheless, we will see that the same charge confinement mechanism described for the Ti 3d conduction states in STO/LAO\cite{delugas} applies here as well, and a 2DEG of few nm extension shows up at the interface.

In absence of polarization catastrophe, mobile charge must be furnished either by field effect or native defects, or by explicit doping of the SZO side (modulation doping). Field-effect is a very efficient way to manipulate the amount of carriers confined at the interface of oxide heterostructures. As such, it is not only powerful mechanism for current-switching applications but also a convenient and clean way to explore the phase diagram of these systems, as demonstrated by recent reports of field-effect modulated 2DEG superconductivity in STO/LAO\cite{reyren,caviglia}, STO\cite{ueno}, and KTaO$_3$\cite{ueno1,yoshi}.  Concering chemical doping, we will explore STO/SZO superattices with Zr-Nb and Sr-La substitutions in a single ZrO$_2$ and SrO layer, respectively, thus mimicking the heterostructures realized in Refs.\onlinecite{ohta_nm} and \onlinecite{jang}. We will see that chemical doping further reduces the extension of 2DEG, which thus become ultra-confined in about 1 nm at the STO/SZO interface. Nicely, our predictions are confirmed by recent experiments\cite{kajdos} published right after the completion of this work, revealing the presence of a 2DEG in STO/Sr(Ti$_{1-x}$Zr$_x$)O$_3$ heterostructures. 

Our results support the viewpoint that the polar catastrophe in STO/LAO, while functional to the charge transfer from surface to interface, is not instrumental to the confimenent process in itself. In fact, essential ingredients for the electron charge confinement are the chemistry of 3d Ti orbitals, the large misalignment between STO and SZO conduction bands, and the electric field due to the electrostatic potential offset at the interface.

\section{Method}

The accurate ab-initio description of oxide heterostructures requires an advanced computational method capable of reproducing the band gap of the individual components as well as their band alignment at the interface, both of which are  generally outside the capability of standard local-density functional (LDA) methods. Here we use the pseudo-self interaction correction approach (VPSIC)\cite{vpsic}, successfully applied to study a number of magnetic and non-magnetic materials\cite{archer} and a series of oxide heterostructures, including the STO/LAO interface\cite{delugas,filippetti}, Nb:STO\cite{delugas1} and LNO/LAO\cite{puggioni} superlattices. This method is an empowered variant of the formerly known pSIC approach\cite{psic,psic1}, successfully applied to a vast range of strong-correlated systems, including manganites\cite{manganiti}, cuprates\cite{cuprati}, high-K materials\cite{high-k}, and diluted magnetic semiconductors\cite{dms}. The VPSIC approach used in this work is implemented in the PWSIC code working on plane-waves plus ultrasoft pseudopotential\cite{uspp} basis set. For these calculations we used well-converged cut-off energies (35 Ryd) and dense k-point grids for the Brillouin-zone integration. 

For bulk STO and SZO the VPSIC-calculated band gap of 2.9 eV and 5.3 eV (against experimental 3.2 eV and 5.6 eV values) enormously improve the typically 50\%-underestimated LDA values; also accurately reproduced are the equilibrium lattice constants calculated for STO (3.915 \AA, against experimental 3.905 \AA) and SZO (4.13 \AA, vs. experimental 4.11 \AA). For the simulation of the interfaces we use supercells of 1$\times$1 and $\sqrt{2}\times\sqrt{2}$ symmetry in the plane, depending on the considered doping concentration, and orthogonal size equal to 13 unit cells (about 50 \AA) for single interfaces and up to 17 unit cells (66 \AA) for the superlattices.

Charge doping is introduced in two ways. First, for small doping concentrations, we simply add external electron charge with a positive homogeneous charge background restoring neutrality; the electronic structure corresponding to the augmented electron charge is recalculated self-consistently at each doping, thus bypassing the faulty rigid band approximation. This approach mimics a field-effect charge manipulation or charge from native donors of the STO substrate. Second, at large doping concentration we explicitly describe chemical doping by Zr$\leftarrow$Nb and Sr$\leftarrow$La atomic substitution.

\section{Results: STO/SZO single interfaces}

\begin{figure}
\centerline{\includegraphics[clip,width=9cm]{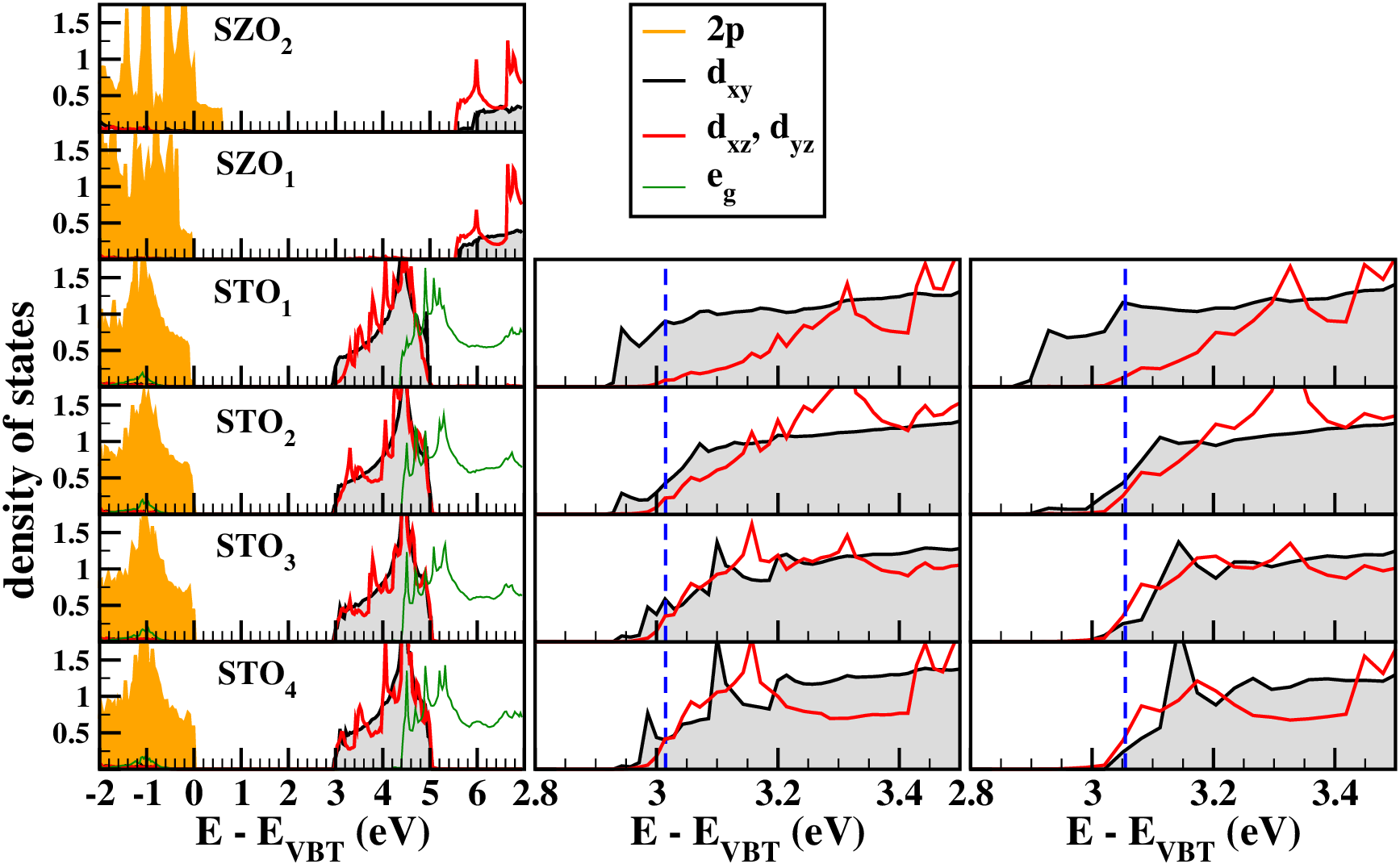}}
\caption{Orbital-projected DOS calculated for the STO$_7$/SZO$_2$ interface. STO$_i$ and SZO$_i$ indicates the i$^{th}$ layer from the interface. Left panels are for the undoped interface, central and right panels for the interface doped with 0.05 and 0.1 electrons per unit area, respectively. For the latter, the DOS is magnified in a small region around E$_F$, where SZO layers do not contribute. Only O 2p, Ti 3d, and Zr 4d DOS are shown, as indicated by the labels. Blue dashed lines indicate E$_F$ (energy zero is fixed at the valence band top).
\label{dos}}
\end{figure}

We start considering a (STO)$_7$/(SZO)$_2$/vacuum stoichiometric supercell mimicking a single interface of 2 layers of SZO grown on the STO substrate. Notice that at variance with STO/LAO, this system has only one type of interface, namely TiO$_2$/SrO/ZrO$_2$. The calculated density of states (DOS) decomposed in orbital contributions  is reported in Fig.\ref{dos}, left panels. For clarity, only the energy region surrounding the band gap is shown. As expected the system is insulating, with a band gap of about 3 eV in the STO side between O 2p valence bands (VB) and Ti 3d conduction bands (CB). In the surface SZO layer, the band gap is between O 2p and Zr 4d states, and slightly reduced in size (4.9 eV) with respect to its bulk value 5.3 eV due to local structural relaxations. The VB and CB offsets (0.5 eV and 2.5 eV respectively) are substantial, and similar to those obtained in STO/LAO. 

Then, exploiting our supercell technique, we introduce excess electron charge, to be interpreted as due to field-effect modulation. We consider two sheet concentrations, n$_{2D}$=3.3 and 6.6$\times$ 10$^{13}$cm$^{-2}$ (0.05 and 0.1 electrons per interface area, respectively). The corresponding DOS, displayed in the middle and right panels of Fig.\ref{dos}, respectively, clearly show that the additional charge does not spread through the STO substrate, but progressively accumulates mainly in the d$_{xy}$ orbital of the Ti layer closest to the interface (STO$_1$, indicated by the filled gray area in the Figure). The onset of the DOS for this orbital is indeed visibly lower in energy than the (d$_{xz}$, d$_{yz}$) DOS sited on the same Ti (STO$_1$, red line). Hereafter we call {\it on-site splitting} the difference between d$_{xy}$ and (d$_{xz}$, d$_{yz}$) conduction band bottoms (CBB), and {\it inter-site splitting}, the difference between CBBs of d$_{xy}$ states sited on different Ti layers (being highly localized in the {\it xy} plane, each d$_{xy}$ can be unambiguously associated to a given layer). 

\begin{figure}
\centerline{\includegraphics[clip,width=8.5cm]{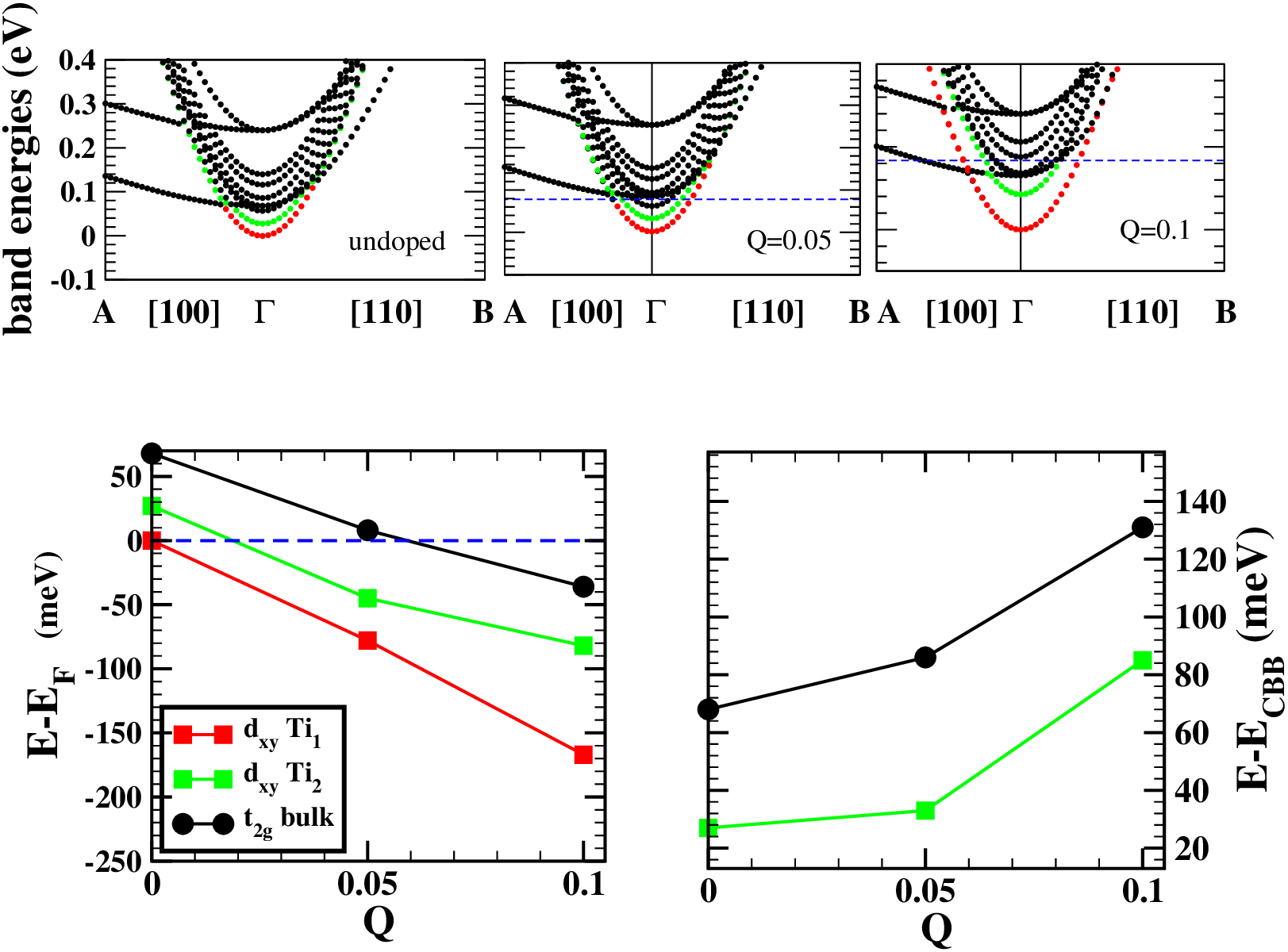}}
\caption{Top: calculated band energies for (left to right panels) the STO$_7$/SZO$_2$ undoped interface, and the same interface with Q=0.05 and 0.1 electrons per unit interface area. Colors correspond to their orbital character labeled in the bottom panel. Blues dashed lines indicate E$_F$. K-point coordinates (unit of 2$\pi$/a) are A=(1/4,0,0), B=(1/4,1/4,0). Bottom left panel: CBB for the two lowest d$_{xy}$ states (Ti$_i$ is the i$^{th}$ Ti from the interface) and the t$_{2g}$ band manifold of bulk-like Ti atoms, calculated with respect to E$_F$ as a function of Q (for Q=0 E$_F$ is fixed at the bottom of the lowest band). Bottom right panel: CBB of Ti$_2$ d$_{xy}$ (green squares) and bulk-like t$_{2g}$ states (black circles) relative to the Ti$_1$ d$_{xy}$ states. They represent inter-site and on-site splitting, respectively (see text).
\label{band}}
\end{figure}

A quantitative analysis of these orbital splittings is given in Fig.\ref{band}, where band structure and band bottom misalignment at varying doping is reported. Remarkably, both the on-site Ti$_1$ splitting (bottom right panel, black circles) and the inter-site Ti$_1$-T$_2$ splitting (bottom right, green squares) are sizable even for the undoped insulating system (65 meV and 25 meV, respectively), and then they progressively grow with the carrier concentration.

These results indicate that 2D charge confinement in STO/SZO (analogously to STO/LAO) is governed by the size of 3d t$_{2g}$ orbital splitting. The t$_{2g}$ on-site splitting originates from the bi-dimensional and very anisotropic nature of Ti t$_{2g}$ orbitals, which results in very different planar (m$^*$=0.7 m$_e$) and orthogonal (m$^*$=8.8 m$_e$) effective masses\cite{delugas}. Since d$_{xz}$ and d$_{yz}$ have lobes oriented orthogonally to the interface, they pay an energy penalty due to the sharp interface potential and shift up in energy with respect to the d$_{xy}$ state -- to put it differently, the hopping between adjacent Zr 4d d$_{xz}$ and d$_{yz}$ orbitals is suppressed by the huge (2.5 eV) Zr-Ti CB offset. Clearly the on-site splitting is maximum for Ti$_1$ but it rapidly disappears away from the interface. 

\begin{figure}
\centerline{\includegraphics[clip,width=8.5cm]{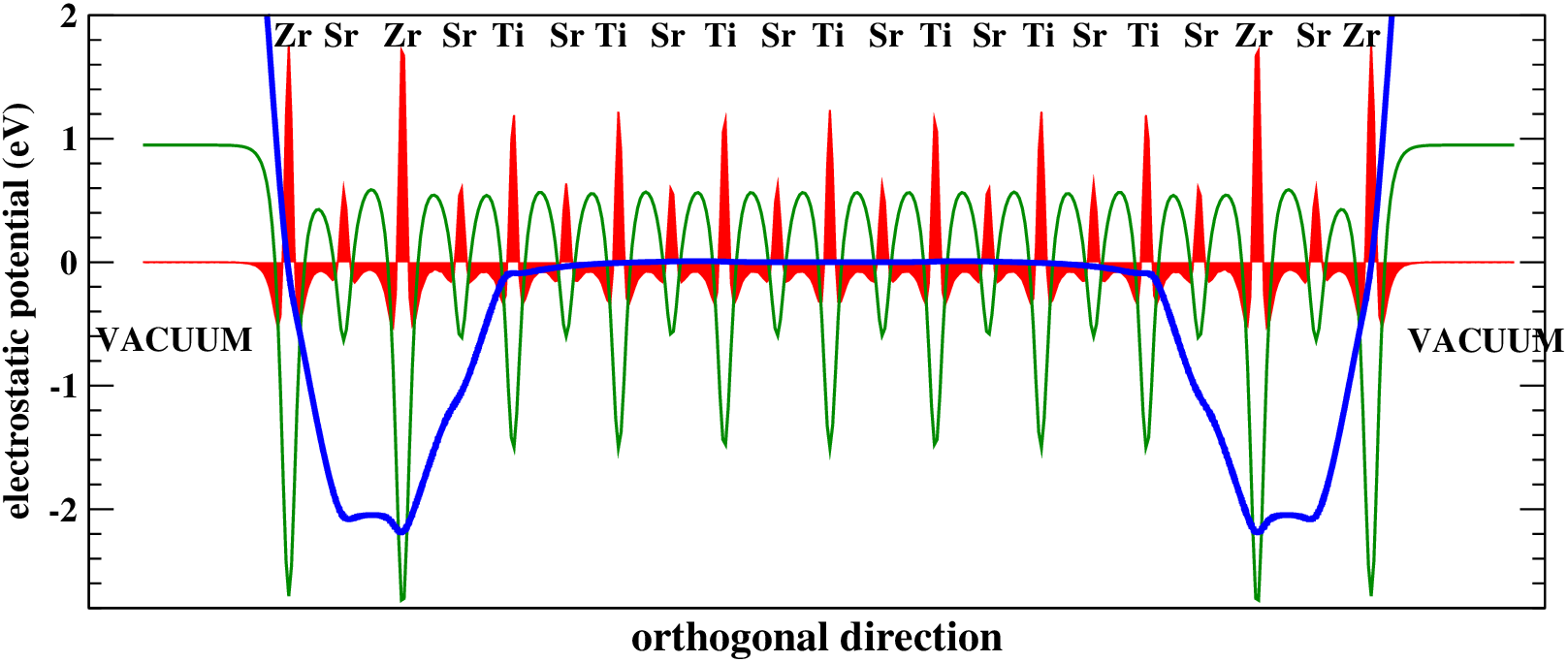}}
\caption{Macroscopic (blue line) and planar (green) average of the electrostatic potential in the direction orthogonal to the interface\cite{macro}. The planar average of the total (electronic plus ionic) charge density (red filled line) is also shown. The positive red peaks indicate the positions of the (100) layers. For clarity, the charge planar average is magnified by a factor 4, the potential planar average reduced by a factor 10.
\label{pot}}
\end{figure}

The interpretation of the d$_{xy}$ inter-site splitting is more subtle. We may expect the STO d$_{xy}$ bands to be almost unaffected by the interface and hence remain substantially bulk-like and degenerate. In turn this would leave the charge homogeneously spread through STO. To shed light on this aspect we calculated the average\cite{macro} electrostatic potential along the stacking direction (Fig.\ref{pot}). As expected, the potential is about constant in the STO side, signaling bulk-like behaviour. However, the large electrostatic potential offset ($\sim$ 2 eV) between the (more repulsive) STO side and the SZO overlayer causes a sharp potential gradient, hence a local electric field $\sim$5 $\times$10$^7$ V/cm acting across the interface region which includes the Ti$_1$ layer. This field causes a $\sim$0.1 eV energy lowering of the Ti$_1$ 3d states with respect to the bulk-like Ti 3d states of the substrate, and drives the 2D charge confinement. The electrostatic misalignment can be ultimately attributed to the difference between Zr and Ti: both are 4+ ions, but the former is less electronegative (1.33 against 1.54 of Ti in Pauling units), thus it leaves more electrons to the surrounding oxygens. Indeed Fig.\ref{pot} shows that the planar-averaged total charge (red line) is higher (more positive) in ZrO$_2$ than in TiO$_2$ layers, and correspondingly the electrostatic potential (green line) has deeper attractive wells for ZrO$_2$ than for TiO$_2$.

\section{Results: STO/SZO superlattices}

Mobile charge can be also introduced at the interface by explicitely doping the SZO layer, e.g. with Zr$\leftarrow$Nb and Sr$\leftarrow$La substitution. In the following we consider STO$_{10}$/SZO$_m$ superlattices (SLs) (with m up to 7) and doping restricted to a single central SZO layer. These structures are similar in spirit to the Nb-doped\cite{ohta_nm,mune,delugas1} and La-doped\cite{jang} STO SLs where a confined 2DEG forms for appropriate doping and layer thickness. Here we show that the large band offset at the STO/SZO interface strongly enhances the charge confinement present in the doped STO SL, and causes the formation of a very narrow 2DEG. 

\begin{figure}
\centerline{\includegraphics[clip,width=9cm]{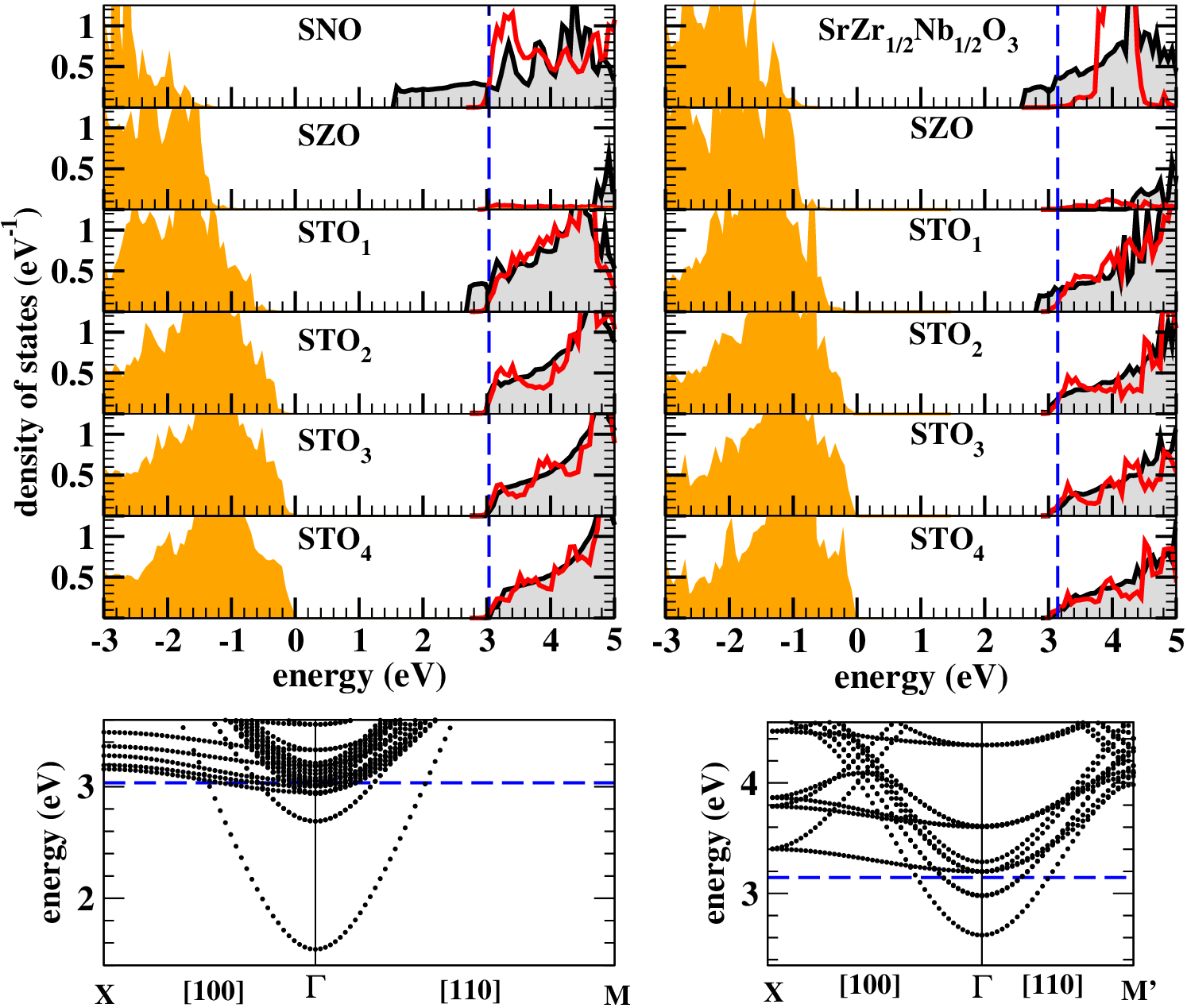}}
\caption{Upper panels: Orbital-resolved DOS for 100\% Nb-doped (left panels) and 50\% Nb-doped (right panels) STO$_{10}$/SZO$_m$ SL with m=3. In both cases, doping (Nb-Zr substitution) is restricted to the single SZO central layer (the uppermost), while the lowest is central to the STO side. Color codes for specific orbital contributions are the same as in Fig.\ref{dos}. Blue dashed lines indicate $\epsilon_F$. Lower panels: band energies corresponding to the DOS illustrated above. K-space coordinates are (in units of 2$\pi$/a) X=(1/2,0,0), M=(1/2,1/2,0), M'=(1/4,1/4,0).
\label{nb_dop}}
\end{figure}

In Fig.\ref{nb_dop} orbital-resolved DOS and band energies for the m=3 SL are shown with 50\% and 100\% Zr$\leftarrow$Nb substitution in the central SZO layer. These two doping concentrations donate 0.25 and 0.5 electrons to each of the two TiO$_2$/SrO/ZrO$_2$ interfaces of the supercell, respectively. According to the band alignment seen for the STO/SZO single interface, we may expect a modulation-doping mechanism which transfers the electron charge from the Nb-doped SZO layer to the undoped STO side. In fact, charge transfer is not complete since Nb$^{5+}$ is a strongly electronegative ion, thus about 50\% of the charge remains trapped in the doped SZO layer, and 25\% transferred to each of the two STO sides. We see from the calculated DOS that at both doping concentrations the SL displays a tightly confined 2DEG, with most of the mobile charge trapped in the planar d$_{xy}$ bands of just three layers: the Nb-doped layer (whose CBB lies $\sim$1.5 eV and 0.54 eV below E$_F$ for 100\% and 50\% doping, respectively) and the two interface-Ti layers (whose CBB is 0.3 eV below E$_F$ for both doping levels). The d$_{xy}$ charges sitting on the Nb-doped layer and the two interface-Ti layers are well separated by the energy barrier due to the undoped SZO layer, whose CBB is placed 1.6 eV above E$_F$. However, we can see a residual tunneling contribution from the orthogonally-extended (d$_{xz}$, d$_{yz}$) states across the barrier (red line in the ZrO$_2$ panels of Fig.\ref{nb_dop}). This tunneling can be suppressed by increasing the SZO thickness. To the aim, we performed calculations for thicker SLs up to m=7 (thus with 3 SZO layers between the Nb-doped layer and the STO sides). Tunneling apart, our results indicate that band alignments and charge distribution remain almost unaffected by the increased SZO thickness, thus one SZO layer is already sufficient to induce a fully confined 2DEG.

\begin{figure}
\centerline{\includegraphics[clip,width=6.5cm]{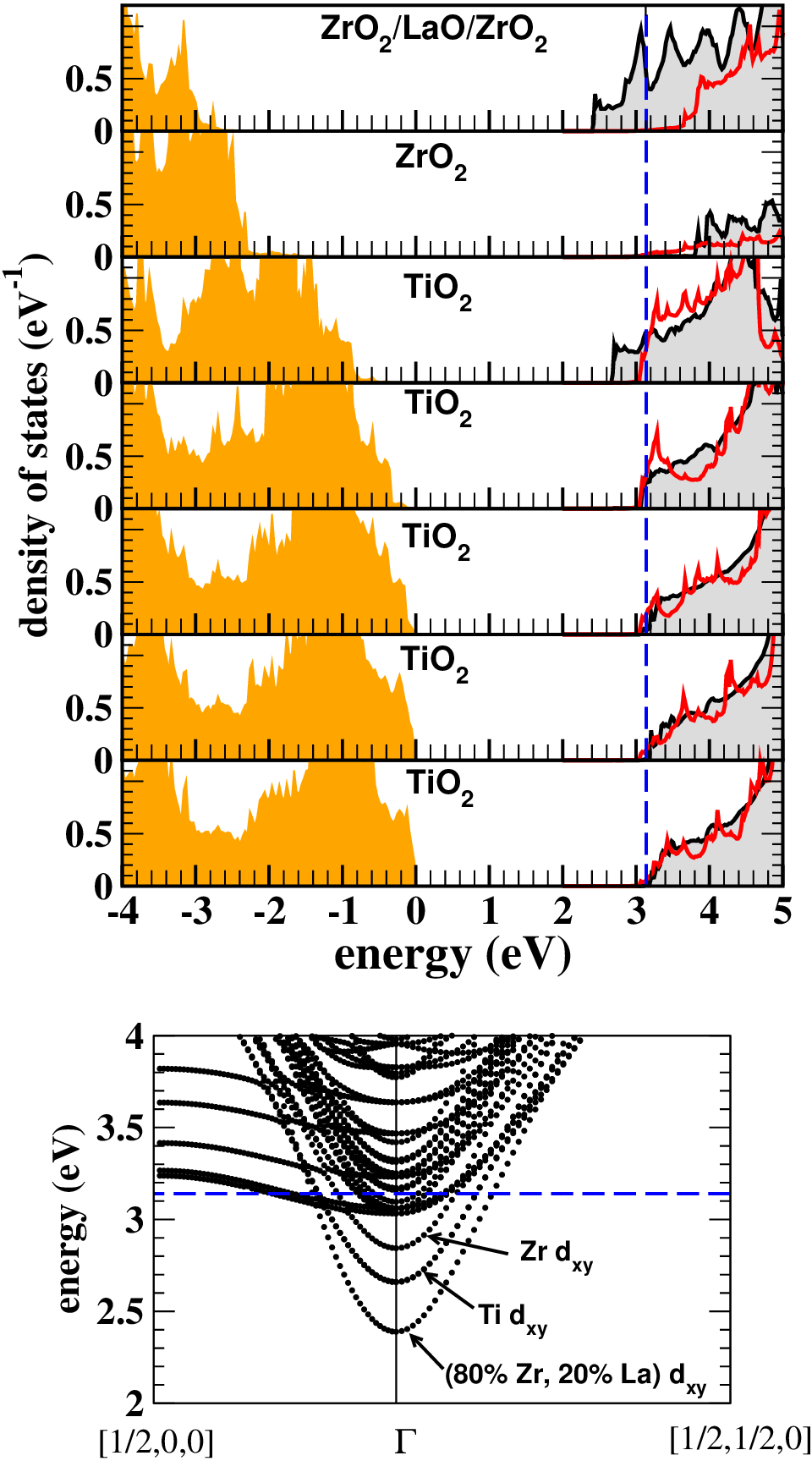}}
\caption{Upper panels: Orbital-resolved DOS for La-doped STO$_{10}$/SZO$_m$ SL with m=4. Doping is included via a Sr-La substitution in the central LaO layer of the SZO$_4$ side. Upmost panel includes the DOS spanning three monolayers: ZrO$_2$/LaO/ZrO$_2$, the lowest is central to the STO side. Color codes for specific orbital contributions are the same as in Fig.\ref{dos}. Blue dashed lines indicate $\epsilon_F$. Lower panels: band energies corresponding to the DOS illustrated above. K-space coordinates are in units of 2$\pi$/a.
\label{la_dop}}
\end{figure}

This tight 2DEG confinement has two causes: the presence of the SZO barrier which suppresses the (d$_{xz}$, d$_{yz}$) occupancy, and the attractive action of the ionized Nb dopants. Concerning the role of (d$_{xz}$, d$_{yz}$) states, it is important to emphasize the difference between this SL and the STO/LAO interface: in the latter a significant portion (about a third) of the 0.5 electrons per unit interface (predicted by the polarization catastrophe) spreads through a multitude of bulk-like (d$_{xz}$, d$_{yz}$) bands, thus resulting in a much broader 2DEG (measured thicknesses\cite{nakagawa,basletic,singh,dubroka, copie, pallecchi} indeed reach over $\sim$10 nm). In the present case, these states are scarcely populated at 100\% doping, and substantially empty at 50\% doping, thus the 2DEG has an almost pure d$_{xy}$ character. Nb dopants also excert a remarkable confining action: comparing the DOS profiles in Figs.\ref{dos} and \ref{nb_dop}, we can see that the valence band top of the Nb-doped SZO layer is downshifted from +0.6 eV of undoped STO/SZO to -1.6 eV (-0.9 eV) for 100\% (50\%) doping, thus amplifying the STO/SZO electrostatic potential offset from $\sim$2 eV of undoped STO/SZO to $\sim$4.2 eV (3.5 eV) for the 100\% (50\%) doped system.

Results for La-doping are illustrated in Fig.\ref{la_dop}, where we report DOS and band energy calculations for a STO$_{10}$/SZO$_2$/LaO/SZO$_2$ SL with a central LaO layer (i.e. 100\% Sr-La substitution) sandwiched within two SZO units on both sides. Even in this case a tightly confined 2DEG shows up, with the charge fully confined within five STO layers from the interface. However, La$^{3+}$ is less electronegative than Nb$^{5+}$, thus the confinement (i.e. about 2 nm thickness) is slightly looser than for 100\% Nb-doping (1 nm). Now, half-electron remains trapped within three monolayers (ZrO$_2$/LaO/ZrO$_2$) of the doped side (only a minor portion in LaO, more than 90\% in two adjacent ZrO$_2$), at variance with the 100\% Nb-doping case, where half-electron was trapped in just a single NbO$_2$ monoloyer. This slightly broader charge distribution results in a reduced La$^{3+}$ impurity screening and a wider STO/SZO valence band offset (2.6 eV). For the same reason, on the other hand, the STO/SZO conduction band offset decreases, resulting in a reduced STO/SZO conduction barrier (0.6 eV agains 1.6 eV of 100\% Nb-doping). The remaining half-electron is distributed in equal parts in the two undoped STO sides. In Fig.\ref{la_dop}, lower panel, we see the presence of three occupied d$_{xy}$ bands; the lowest (0.75 eV below E$_{F}$) and third-lowest (0.3 eV below E$_F$) result from a bonding-antibonding splitting involving the two occupied Zr d$_{xy}$ orbitals and, on a minor extent, the La d$_{xy}$ orbital; the second-lowest band (0.5 eV below E$_F$) is double-degenerate, and related to the two interface-Ti layers. Due to the smaller conduction barrier, for La-doping there is more tunneling across the SZO barrier and a larger fraction of interface-Ti d$_{xz}$, d$_{yz}$ charge than for Nb-doping.

\section{Summary and Conclusions}

In summary, we have described the presence of a tightly confined electron gas in STO/SZO heterostructures, configured either as single interface or doped SL. For the single interface, mobile charge was introduced in order to mimick field-effect or remote native defects. For SLs field-effect typically does not work, thus we considered chemical doping in form of Nb-Zr and La-Sr substitutions. 

In absence of chemical doping, we found that the 2D confinement is produced by a combination of highly localized 3d orbitals, large conduction band offset, and the built-in electric field in the narrow interface region. The charge introduced by field effect accumulates in a STO region near the interface of $\sim$2-3 nm thickness, thus with characteristics similar to STO/LAO. If Nb-doping or La-doping is added to SZO, the 2DEG becomes extremely narrow (1 nm or less for Nb-doping, 1-2 nm for La-doping), and can be essentially described as a two-band system: one Zr-Nb (or Zr-La) t$_{2g}$ d$_{xy}$ band which collects the charge in the doped region, and one Ti t$_{2g}$ d$_{xy}$ band (per interface) hosting the charge reversed in the undoped STO side (modulation doping). Due to the large STO/SZO conduction band mismatch, even a single undoped SZO unit cell interposed between STO and the doped SZO region is sufficient to suppress orthogonal hopping and in turn the occupancy of the orthogonally-oriented (d$_{xz}$, d$_{yz}$) states (substantially empty for Nb-doping, marginally occupied for La-doping). We remark that this occupancy is substantial in STO/LAO\cite{delugas} and in Nb-doped STO SLs \cite{delugas1} and contributes to a much broader 2DEG in these systems.

This almost pure d$_{xy}$ electron gas in STO/SZO is expected to have a larger in-plane mobility than an equivalent 2DEG with partial d$_{xz}$, d$_{yz}$ orbital contributions, given the smaller effective mass of the former. Furthermore, narrow confinement is a paramount quality for the 2DEG field-effect functionalities: on the one hand, gate fields (V$_G$) cannot penetrate deeper than a Thomas-Fermi screening length (for a metal this can be only a few nm); on the other hand, maximum V$_G$'s  are limited to about 10 MV/cm to avoid current leakage or voltage breakdown, and so is the charge density that can be modulated: Q= en$_{3D}$AL=V$_G$(A$\epsilon$/L$_d$), where A and L$_d$ are area and thickness of the dielectric, and L the 2DEG thickness. Clearly, the smaller the 2DEG thickness L, the larger the charge density potentially switchable by the electric field.

Our results are consistent with recent experiments for the STO/Sr(Ti$_{1-x}$,Zr$_x$)O$_3$ (STO/STZO) interface grown by molecular beam epitaxy\cite{kajdos}. The actual system considered in Ref.\onlinecite{kajdos} is a single interface between very thick STO and STZO sides with $x$=0.05 Zr content $x$ (our calculation assumes $x$=1), with STZO doped by La inclusion. The presence of a 2DEG in the STO side is demonstrated by 2D Shubnikov de-Haas resistivity oscillations, and the amount of 2D charge quantified in about 19\% of the total excess charge introduced in the sample by La-doping, with the remaining portion supposed to stay in the STZO side. The modulation-doping scenario depicted in this work is closely analogous to our La-doped STO/SZO superlattice. 

A final consideration concerns the practical realization of the ideal STO/SZO heterostructures considered in this work. Due to the large (5\%) planar mismatch, we may expect that above a certain SZO critical thickness, large planar strain may cause cracks, dislocations and stacking faults, in turn causing the disruption of the 2D mobility of the system. Considering ultrathin SZO film is a way to circumvent this problem. We expect few SZO layers to adapt epitaxially to the STO substrate and grow without disorder. A reference point is STO/LAO where, despite the 3\% mismatch, LAO is grown perfectly epitaxial for up to about 15 unit cells. A similar (or a bit lower) critical thickness should be expected for STO/SZO interface and SLs as well.

\acknowledgments

Work supported in part by MIUR-PRIN 2010 {\it Oxide}, IIT-Seed NEWDFESCM, IIT-SEED POLYPHEMO and ``platform computation" of IIT, 5 MiSE-CNR, Fondazione Banco di Sardegna and CINECA grants, and Cybersar Cagliari.



\end{document}